\documentclass{article}

\usepackage{PRIMEarxiv}

\usepackage[utf8]{inputenc} 
\usepackage[T1]{fontenc}    
\usepackage{hyperref}       
\usepackage{url}            
\usepackage{booktabs}       
\usepackage{amsfonts}       
\usepackage{nicefrac}       
\usepackage{microtype}      
\usepackage{lipsum}
\usepackage{fancyhdr}       
\usepackage{graphicx}       
\graphicspath{{media/}}     
\usepackage{amsmath}
\usepackage{multirow}

\usepackage{graphicx}
\usepackage{array}
\usepackage{colortbl}
\usepackage{xcolor}
\xdefinecolor{gray75}{gray}{0.85}

\title{Assessing the Impact of Mobile Attackers on RPL-based Internet of Things}

\author{
  Cansu Dogan \\
  WISE Lab., Department of Computer Engineering \\
  Hacettepe University \\
  Ankara\\
  \texttt{cansu-dogan@hacettepe.edu.tr} \\
   \And
  Selim Yilmaz \\
  WISE Lab., Department of Software Engineering\\
  Mugla Sıtkı Koçman University\\
  Mugla \\ 
  \texttt{selimyilmaz@mu.edu.tr} \\
  \And
  Sevil Sen \\
  WISE Lab., Department of Computer Engineering \\
  Hacettepe University \\
  Ankara\\ 
  \texttt{ssen@cs.hacettepe.edu.tr} \\
}

\begin{document}
\maketitle

\begin{abstract}
The Internet of Things (IoT) is becoming ubiquitous in our daily life. IoT networks that are made up of devices low power, low memory, and low computing capability appears in many applications such as healthcare, home, agriculture. IPv6 Routing Protocol for Low Power and Lossy Network (RPL) has become a standardized routing protocol for such low-power and lossy networks in IoT. RPL establishes the best routes between devices according to the requirements of the application, which is achieved by the Objective Function (OF).  Even though some security mechanisms are defined for external attackers in its RFC, RPL is vulnerable to attacks coming from inside. Moreover, the same attacks could has different impacts on networks with different OFs. Therefore, an analysis of such attacks becomes important in order to develop suitable security solutions for RPL. This study analyze RPL-specific attacks on networks using RPL's default OFs, namely Objective Function Zero (OF0) and the Minimum Rank with Hysteresis Objective Function (MRHOF). Moreover, mobile attackers could affect more nodes in a network due to their mobility. While the security solutions proposed in the literature assume that the network is static, this study takes into account mobile attackers.
\end{abstract}


\section{Introduction}
\label{sec:introduction}
IoT has become one of the most revolutionary concepts of this century with the advancements in sensor and networking technologies. The adoption of the IPv6 protocol standard has led to the emergence of numerous IoT devices that are capable of communicating with each other and with remote machines through the Internet. It is estimated that total installed base of IoT devices will reach around 75 billion in 2025~\cite{Statistica}. IP-connected IoT devices have recently opened the door to the development of several life-enhancing applications. These include healthcare monitoring, smart cities, transportation and logistics, military and defense, robots, and the like.

Low Power and Lossy Networks (LLN) in IoT are characterized by high packet loss and low throughput. Due to the characteristics of LLN, traditional routing protocols, even those proposed for WSNs are not applicable to LLNs. Therefore, the Internet Engineering Task Force-Routing over LLN (IETF-RoLL) group designed an IPv6-based routing protocol specific for the LLNs: RPL, which operates on the IEEE 802.15.4 standard using IPv6 over Low-power Wireless Personal Area Network (6LoWPAN) adaptation layer. 

RPL makes use of OF in order to build an optimal route between the IoT devices (or nodes) in the network. There are different routing metrics employed in OFs that play key role in selecting parent node and hence routes to the root node or destination nodes. These include Expected Transmission Count (ETX), hop count, energy, and the like~\cite{of_metrics}. There is no obligation to use a specific OF metric; it often depends on the requirements of IoT applications. The appropriate selection of OFs is very important because it significantly affects the performance of the network including packet delivery ratio, end-to-end delay, power consumption. Although various kinds of OFs have been proposed until far, OF0 and MRHOF are known as standard OFs defined for the RPL.

RPL ensures efficient routing among IoT devices on LLN, and that's why, it is adopted as the standard routing protocol today. However, there are a number of significant challenges faced by RPL. The first is that the RPL protocol is vulnerable to attacks (particularly to insider attacks) that aim to consume resources of IoT devices and hence reduce the lifetime of the network. The other is that RPL does not support mobility, it is specifically designed for static networks. Considering the fact that most of the application scenarios in IoT such industrial automation involve the use of mobile nodes attached to agents such as workers, robots, products, and the like, this can be regarded as one of the major drawbacks of RPL. Therefore, new improvements on RPL have been explored by researchers~\cite{kamgueu2018survey,lamaazi2018rpl}. Such improvements should be carried out for providing its security as well, since the existence of mobile attackers could severely damage the network. This is one of the main objectives of the current study.

This study explores the performance of RPL under attack. Although some analysis is carried out for a particular type of attack, such as rank attack~\cite{le2013impact}, version number attack~\cite{mayzaud2014study,aris2016rpl}; this study differs from these studies by taking into account both OF and mobility that could change the effect of attacks on networks. Moreover, this study does not focus on a particular attack but covers different types of attack namely version number, DIS flooding, worst parent attacks are included in this study. As stressed earlier, the mobility of the nodes can harm the RPL-operated network with changing rates depending on several factors. Among them, the OF used in the network and the density of mobile nodes are priori because they are highly correlated to the performance of RPL. This becomes alarming when there are attack nodes present in the network. From this point of view, this study will provide a great insight to the researcher studying to enhance RPL towards the mobility and to develop security solutions. This is the continuation study of~\cite{Doan2022AnalysisOR} which analyze attacks on static networks only. To the best of our knowledge, this is the first study that thoroughly analyzes the performance of RPL on networks with varying mobile attacker densities and with different OFs. RPL is analyzed under different network scenarios by using the following performance metrics: packet delivery ratio, power consumption, overhead, and latency. 

The rest of the paper is organized as follows. The background information that covers an overview of RPL, the standard/default objective functions used in RPL, and attacks targeting RPL is given in Section~\ref{section_background}. Section~\ref{section_related_works} summarizes the studies in the literature that analyze attacks and consider mobility in RPL security. The experimental settings are introduced, and the experimental results are discussed in Section~\ref{section_experiment}. Finally,  Section~\ref{section_conclusion} concludes the study.

\section{Background}
\label{section_background}
\subsection{Overview of RPL}

RPL creates a topology called Destination Destination Oriented Directed Acyclic Graph (DODAG). DODAG is a DAG rooted at single destination.
A network can operate on one or more RPL instances where multiple DODAGs can take part. The role of each instance is to define an objective function to calculate the optimum path within the DODAG. A DODAG is built by using the following RPL control packets:
\begin{itemize}
	\item \textit{DODAG Information Object (DIO):} It is initiated and broadcast only by the root node. DIO packets carry network information (e.g., instance ID, version number). Each of the receiving nodes adds the sender to its parent list, calculates its own rank value, which states its position in the graph with respect to the root node, and finally, it forwards DIO to its neighbors. DIO packets are relayed throughout the graph and play a major role in constructing the default upward routes. The transmission interval of the DIO packets should be well adjusted. The lower the interval is the higher the overhead leading to shorter lifetime of the network; the higher the interval is, however, the lower the responsiveness to the network's inconsistencies. The management of transmission rate of DIO packets by the nodes is governed by an algorithm in RPL called \textit{trickle timer}. This algorithm enables DIO packets being broadcast more frequently initially to make DODAG stable, and increases the time interval to avoid unnecessary propagation of DIO packets in the network. The timer is reset when an inconsistency is reported by the nodes, causing the DIO packets being broadcast instantly again.
	\item \textit{DODAG Information Solicitation (DIS):} It is used as a solicitation for having DIO information when a new node is to join the DODAG. DIS packets are broadcast by the new node to its neighbors.
	\item \textit{Destination Advertisement Object (DAO):} It is used for the construction of the downward routes from the root to sensor nodes. Based on the mode of operation, 
	the child unicasts DAO packets either to root node in non-storing mode or to its selected parent node in storing mode so that it records downward routes in its routing table for the sub-DODAG.
	\item \textit{Destination Advertisement Object Acknowledgement (DAO-ACK):} Upon receiving DAO packets from a parent node, DAO-ACK packets are sent to the the sender node as an acknowledgement. 
\end{itemize}

\subsection{Objective functions in RPL}
RPL objective function is used for the calculation of rank value assigned to each node in the network. Therefore, it implicitly governs the selection of the preferred parent of the nodes in the network, and consequently, determines the routing path that is optimal with respect to the utilized OF. The packets are forwarded in the selected routes according to three traffic patterns: point-to-point, point-to-multipoint, and multipoint-to-point.
Objective functions differ with respect to the RPL instances; hence, different OFs could be simultaneously used within an RPL network by different instances. For example, one can take `hop count' into consideration to build routes of a DODAG, the `residual energy' of the nodes can be used for finding the routes of another DODAG in the same network. The selection of appropriate objective functions is critical and changes in accordance with the requirements of the application.

Even though there have been a number of OFs proposed in the literature so far, OF0~\cite{thubert2012objective} and MRHOF~\cite{gnawali2012minimum} are proposed as the default OFs in RPL:

\subsubsection{OF0}
OF0 takes the hop count between the root node and a sensor node into account for the calculation of the rank value of that node. Therefore, it aims to minimize the number of hops to reach to the root node by choosing the node that has the lowest rank from its reachable neighbors as its parent. When OF0 is used as the objective function in the network, for a given node $n$, the rank of this node can be calculated using~\eqref{eq1}.

\begin{equation}
	R(n)=R(p)+RI\label{eq1}
\end{equation}
$R(n)$ is the new rank of node $n$, $R(p)$ is the rank of the preferred parent node, and $RI$ stands for the rank increase metric that is calculated by using~\eqref{eq2}.

\begin{equation}
	RI=\left(Rf \times Sp + Sr\right)\times MHRI \label{eq2}
\end{equation}
$Rf$ is a configurable rank factor and it uses 1 as the default value. $Sp$ is the step of the rank, and $Sr$ is the maximum value assigned to the rank level. $MHRI$ stands for $MinHopRankIncrease$ which is a constant value defined as 256 in RFC6550~\cite{winter2012rpl}.

\subsubsection{MRHOF}
Unlike to OF0, a number of link- and node-based additive routing metrics can be easily integrated into MRHOF. The rank value ($R(n)$ in Eq.~\ref{eq1}), and hence the routing path, is determined according to the employed routing metric which is stored in the `metric container' sub-option in the DIO packet. 

By using one of these routing metrics (e.g., latency, RSSI, and etc.), MRHOF ensures the lowest-cost path in the LLN. Two metrics are integrated into MRHOF in this study: MRHOF with ETX (MRHOF-ETX), which is a link-based metric, and MRHOF with energy (MRHOF-ENERGY), which is a node-based metric. 

MRHOF-ETX chooses the paths with the lowest number of transmission value by considering ETX values of the links. The ETX value of the links is calculated using~\eqref{eq3}. 

\begin{equation}
	ETX=\frac{1}{Df \times Dr}\label{eq3}
\end{equation}
$Df$ is the probability that the neighbor will reach the packet and $Dr$ is the probability that the acknowledgment packet will be received. 

MRHOF-ENERGY chooses the path that provides the maximum remaining energy for the RPL nodes. Energy metric of the nodes is calculated using~\eqref{eq5}.

\begin{equation}
	ENERGY=\frac{P_{max}}{P_{now}}\label{eq5}
\end{equation}
where $P_{max}$ is defined as the targeted maximum power, and it is calculated from the initial energy of node divided by the targeted lifetime; $P_{now}$, however, is the actual power of node.
\subsection{RPL specific attacks}
RPL attacks can vary according to what they primarily target, and they are categorized accordingly: attacks targeting network resources, network topology, and network traffic~\cite{mayzaud2016taxonomy}.  The basic goal of network resource attacks is to use the resources of legitimate nodes and/or the network, resulting in poor network performance. This type of attack can speed the consumption of a node's battery power, use node memory, and cause a delay in the remaining necessary processes. DIS flooding and version number attacks are in the class of attacks that target network resources. RPL attacks can also be used to disrupt the network's topology and worst parent attack is one of the this type of attack. Lastly, there are attacks aimed to disrupt network communication and this category's major goal is to direct network traffic to a specified node. In this study, we have studied version number, DIS flooding, and worst parent attacks.

\begin{itemize}
	\item \textit{Version Number:} Version number in DIO packets is used by the DODAG root to perform global repair, and it is increased only by the root node. In the attack scenario, the malicious node illegitimately increases the incoming version number, causing unnecessary rebuilding of DODAG. In our attack scenario, a malicious node illegitimately increases version number by one in every minute, before forwarding incoming DIO messages to its neighbors.
	\item \textit{DIS Flooding:} It is a typical RPL-specific DoS attack that targets consuming network resources. In order to make nodes or links unavailable in LLNs, attacker continuously sends large amount of control packets. This attack is often performed by sending 
	DIS packets after receiving a DIO packet from a node. By doing so, the DIS flooding attack brings about network congestion and overloading of RPL nodes. In our attack scenario, malicious node multicasts the DIS message to its neighbor nodes every 500 milliseconds.
	\item \textit{Worst Parent:} As stated earlier, an RPL node chooses its own parent node according to the rank value determined by the objective function, which ensures the `best parent' for that RPL node. However, in this attack scenario, attacker node contrarily chooses the worst parent, resulting in non-optimized routing path and hence leading LLN to show very poor performance.
\end{itemize}

\section{Related Works}
\label{section_related_works}
The advancements in the sensor- and actuator-equipped devices and communication technologies in the wireless medium have given rise to the emergence of a great number of IoT applications in recent years. Therefore, IoT-based challenges have become one of the key directions studied by the researchers. The attention towards RPL has been tremendously growing; and until now, a good deal of studies have been proposed on this protocol in the literature \cite{pancaroglu2021load}. 

Although IoT devices are attached to mobile agents (e.g., people, robots, etc.) in many real-world applications, mobility is not supported in the default implementation of RPL. Therefore, although rare, there has been recent attempt that \textit{i}) analyzes RPL on networks under mobility, \textit{ii}) enhances the protocol to integrate the mobility support, \textit{iii}) proposes mobility-aware security solutions, and that \textit{iv}) analyzes RPL against routing attacks. These studies are briefly discussed here. 

There have been few attempts to investigate the performance of RPL-OFs in the static environment; however, we do not take these evaluations into consideration here. Please refer to~\cite{Doan2022AnalysisOR} for the review of these studies. Rather, we focus more on studies that explore the performance of RPL in mobile environments. In~\cite{Zaatouri_et_al}, an evaluation of PRL routing performance is studied in terms of packet loss and power consumption as a function of data traffic density (that is, 6 and 12 packets per minute). The findings reveal that RPL yields higher packet loss and lower power consumption as data traffic increases. Considering the fact that using at least 25 nodes in network simulations is necessary to see multi-hop characteristics of RPL~\cite{kim2017challenging}, the reliability of this evaluation is highly questionable because the experiments are conducted with only two network simulations where only 13 nodes are used of which one is mobile. Another performance evaluation of RPL as a function of two radio duty cycles (that is, 8 and 32 Hz) is made in terms of the packet delivery ratio, ETX, power consumption, and latency in~\cite{Cotrim_et_al}. The duty cycle mechanism is used to listen for a packet transmission of neighbors. To receive incoming packet, the node's radio is turned on when a packet is detected. After receiving, an acknowledgment is sent to the transmitter. A sender node sends its packets during the wake-up period until it receives an acknowledgment. The evaluations suggest that the performance of RPL downs when mobile nodes are present in the network, and the amount of degradation depends on the chosen setting of radio cycle.

In order to cope with downsides caused by mobile nodes in RPL, a great deal of modifications is proposed toward the protocol in the literature. \textit{Trickle algorithm} designed for RPL is not suitable for mobile environments, since it triggers the control packets to be sent periodically by the nodes. Moreover, if the network is stable, the length of time period increases to reduce the overhead in the network. However, a mobile node will instantly need the control packet to connect to the DODAG and to update its preferred parent. This issue is often handled by most of the works to make RPLs adapted to mobile networks. In~\cite{RPL_under_mobility}, the applicability of RPL to Vehicular Ad Hoc Network (VANET) is explored. In order to achieve that, RPL's trickle timer algorithm is disabled because of the vehicles that are highly mobile and often need DIO packets. Instead, a fixed sending time interval is adopted. Therefore, DIO messages are guaranteed to be sent once within each time period, ensuring mobile nodes are connected to DODAG. In addition, ETX values are considered in this approach for the parent selection of mobile nodes. 

Instead of completely disabling the trickle algorithm, there are some approaches that modify the algorithm in the literature.  In~\cite{reverse_trickle_timer_1} a reverse trickle algorithm is proposed. Here, the router nodes set the interval of the trickle timer to a maximum value believing that they remain connected to their parent for a long time. Then, they periodically decrease the interval until it reaches the minimum value. When the time interval reaches the minimum value, children nodes are expected to send DAO packets to control if mobile nodes are connected.  Some assumptions are made for this approach to be applicable to RPL. Firstly, mobile nodes are restricted to be only leaf nodes in DODAG and do not advertise DIO packets. Secondly, mobile nodes are determined with a mobility flag added as a field to the DAO packet. The main drawback of this approach is that an intruder can easily evade this approach once he falsifies the parent nodes by illegitimately changing the mobility flag, which leads to unnecessary propagation of DIO packets. In addition, there is always a static node in range of any mobile nodes. Another approach that modifies the trickle algorithm is proposed in~\cite{reverse_trickle_timer_2}. Here, Received Signal Strength Indicator (RSSI) values are used to configure the time interval of the trickle algorithm. Upon receiving a packet, the nodes read RSSI values, compare them with the last reading values, and send DIS packets immediately after a worsening on RSSI value is detected so that they keep connected to DODAG. 

An adaptive strategy for setting the time interval is integrated into the trickle algorithm in~\cite{Murali_and_Jamalipour}. In the proposed strategy, each mobile node checks the number of neighbors that plays a key role in setting the interval. The more neighboring nodes are, the larger the interval is. In addition to that, the parent selection mechanism is also improved. The rank values are first considered for building the parent list and then the parent node is chosen with respect to, in order, ETX, expected lifetime (ELT), and RSSI. Another adaptive adjustment of time interval is proposed in~\cite{Co-rpl} as an alternative to the trickle timer. It relies on fixed length of time interval which should be adjusted according to the speed of the mobile nodes. This approach enhances RPL with the Corona architecture, a simple concept that divides the network area into the coronas, to locate mobile nodes in the network. Therefore, when links to the parent nodes are broken, mobile nodes are allowed to set the best neighbor node as their preferred parent within the same Corona where the radius is centered at the DAG root. This approach introduces additional flags to DIS and DIO packets. The illegitimate setting of these flags by attackers prevents this approach from functioning properly.

Recently, few security solutions taking into account mobility are proposed. In~\cite{Thulasiraman_and_Wang}, a trust-based security mechanism, which involves a modified rank computation, is proposed against Sybil and DoS attacks. In this approach, not only the OF value but also trust and RSSI values are used for calculation the rank values of nodes. The trust value is used to identify the malicious nodes, and hence, to isolate them. It increases every time a trusted event occurs and decreases when an attempt is malicious. RSSI value, however, is used by mobile nodes to select parent nodes considering the signal strength. Sybil attack is also targeted in~\cite{Murali_and_Jamalipour_ABC} in a network where the mobility-aware RPL proposed in~\cite{Murali_and_Jamalipour} is adopted. Here, Artificial Bee Colony (ABC) algorithm~\cite{ABC} is used to model the behavior of Sybil intruders and to ensure a very harsh attack condition. Then, a lightweight intrusion detection approach is proposed against this attack environment learned. This approach is based on three trust factors driven by \textit{i}) DODAG and NONCE IDs, \textit{ii}) control message counters, and \textit{iii}) timestamps for control messages. Another lightweight security mechanism against the DIS flooding attack, called Secure-RPL, is proposed in~\cite{verma2020mitigation}. Secure-RPL prevents the nodes from resetting the trickle timer redundantly. Therefore, it dramatically reduces unnecessary transmission of DIO packets. 

There are also studies in the literature that analyze the extent to which RPL is affected. when attackers are in the network. In~\cite{mayzaud2014study}, the effect of version number attack is studied as a function of attacker location. The findings reveal that RPL is very vulnerable to this attack, reducing PDR while increasing end-to-end delay and energy consumption of the nodes. In addition, it is found that the more distant the attackers are from the root, the higher the network performance they lead to. The main drawback of this study is that only 20 nodes are considered in the topology of which only one is the attacker. Additionally, the impact of mobility is not studied. This attack is also analyzed in~\cite{aris2016rpl} with respect to two parameters: the initial location of mobile attackers (with respect to the root node) and attacking probability. They found that the initial location has a clear effect on PDR and overhead. As expected, all performance metrics dramatically decrease as the attacking probability increases. The downside of this study is that a single attacker is used at a time, and impact of attacker density is missing. The rank attack is analyzed with different attacker locations in~\cite{le2013impact}. The analysis shows that the bigger the forwarding load area, which is the sum of the forwarding load of all nodes in the area, is, the more impact attack leads to on the network performance. In addition, the cooperation of multiple attackers gives severe damage to the network performance.

As seen, there has been considerable amount of efforts made on RPL because the performance of an RPL-based IoT network is very sensitive to mobility and routing attacks. The degree to which RPL is sensitive to these factors depends on the attack itself, the density of mobile attackers, and the OF used. That's why, it is worth analyzing the performance of RPL with different OFs and attacker densities against various routing attacks in mobile environment, which is the main contribution of this study.
\section{Analysis of RPL Objective Functions Under Mobile Attacks}
\label{section_experiment}

Mobility is a non-trivial concept in cyber security. On the one hand, if the victim nodes are static, attackers due to their mobility could target more victim nodes and hence expand their effect on the network.  Moreover, it allows attackers to evade security solutions. On the other hand, it might limit its effects on mobile victim nodes. Here, we aim to investigate the effect of mobility from the attacker point of view. In the future, we plan to investigate how the mobility of victim nodes limits their exposure to attacks. 

\subsection{Simulation Settings}
In this study, we investigate the behavior of RPL on networks where different number of mobile attackers are involved in networks using different OFs. As explained earlier, we here extend our analysis study proposed in~\cite{Doan2022AnalysisOR} by including mobile attackers into networks and analyzing their effects. In order to ensure a fair comparison between the static and mobile environments, the same networks and simulation settings used in~\cite{Doan2022AnalysisOR} are used here. We adopt \textit{random walk mobility model} for mobile attacker nodes such that they move with a speed of 5 km/h from the beginning of the simulation to the end. We use the Bonnmotion tool~\cite{aschenbruck2013mobility} to generate a realistic movement pattern for mobile attacker nodes. Cooja simulator~\cite{osterlind2006cross} running on Contiki OS (version 2.7)~\cite{contiki-ng} is used to simulate the networks. Each simulation scenario is run with the parameter values listed in Table~\ref{tab1}.

\begin{table}
	\centering
	\caption{Simulation Parameters}
	\begin{tabular}{ll}
		\hline
		\textbf{Simulation Parameters}&\textbf{Values} \\
		\hline Radio Environment & UDGM: Distance Loss\\
		Objective Functions & MRHOF-ETX, MRHOF-ENERGY, and OF0\\
		TX Range & 50m \\
		INT Range & 100m \\
		Simulation Time & 1 hour \\
		Area of Deployment & 200x200 \\
		Number of Sink Node	& 1 node \\
		Number of Sensor Node & 50 nodes\\
		Platform & Sky mote \\
		Traffic Pattern & UDP packets, every 60 sec. by sensor nodes \\
		Mobility Type & Random Walk\\
		Mobile Attacker Speed & 5 km/h\\
		\hline
	\end{tabular}
	\label{tab1}
\end{table}

\subsection{Considered Evaluation Metrics}
\label{section_performance_metrics}
In order to measure the performance of RPL under attacks, the following four performance evaluation metrics are used: \textit{packet delivery ratio}, \textit{overhead}, \textit{power consumption}, and \textit{latency}.

\begin{itemize}
	\item \textit{Packet Delivery Ratio (PDR):} It is calculated from the total number of packets received by the root node divided by the total number of packets sent to the root node. 
	\item \textit{Power Consumption (PC):} In this study, we use the average power consumption of nodes evaluated in mW. Powertrace tool~\cite{dunkels2011powertrace} is integrated into Contiki OS to obtain instant power states of IoT devices. 
	\item \textit{Overhead (OVR):} It is the total number of control messages transmitted by the nodes to create the DODAG. So, overhead is the summation of DIO, DIS, and DAO packets. 
	\item \textit{Latency (LT):} 
	Here, we use the average latency (in seconds) of all packets, which is the time taken from sending to receiving. Packets that are lost or dropped are excluded from the calculation. 

\end{itemize}
\subsection{Simulation Results}
\label{section_results}
In this study, we take the results from~\cite{Doan2022AnalysisOR} as the baseline performance of RPL in static environments.  In addition, these results are used to highlight differences in the performance of RPL when it operates on a network with mobile attackers. For a fair comparison, we here run simulations as done in ~\cite{Doan2022AnalysisOR}, only making attacker nodes mobile. Therefore, the ten networks adopted in~\cite{Doan2022AnalysisOR} are also used in the experiments. In order to explore how large the performance of RPL changes according to the attacker density; one, three, and five nodes are set as mobile attacker nodes for each simulated network, which corresponds to 2\%, 6\%, and 10\%, of all nodes in the network, respectively. It is worth stressing here that the nodes that were used as attacker nodes in~\cite{Doan2022AnalysisOR} are selected as mobile attacker nodes to ensure a reliable comparison in the experiments. 

\begin{figure}[!ht]
    \centering
    \includegraphics[width=\linewidth]{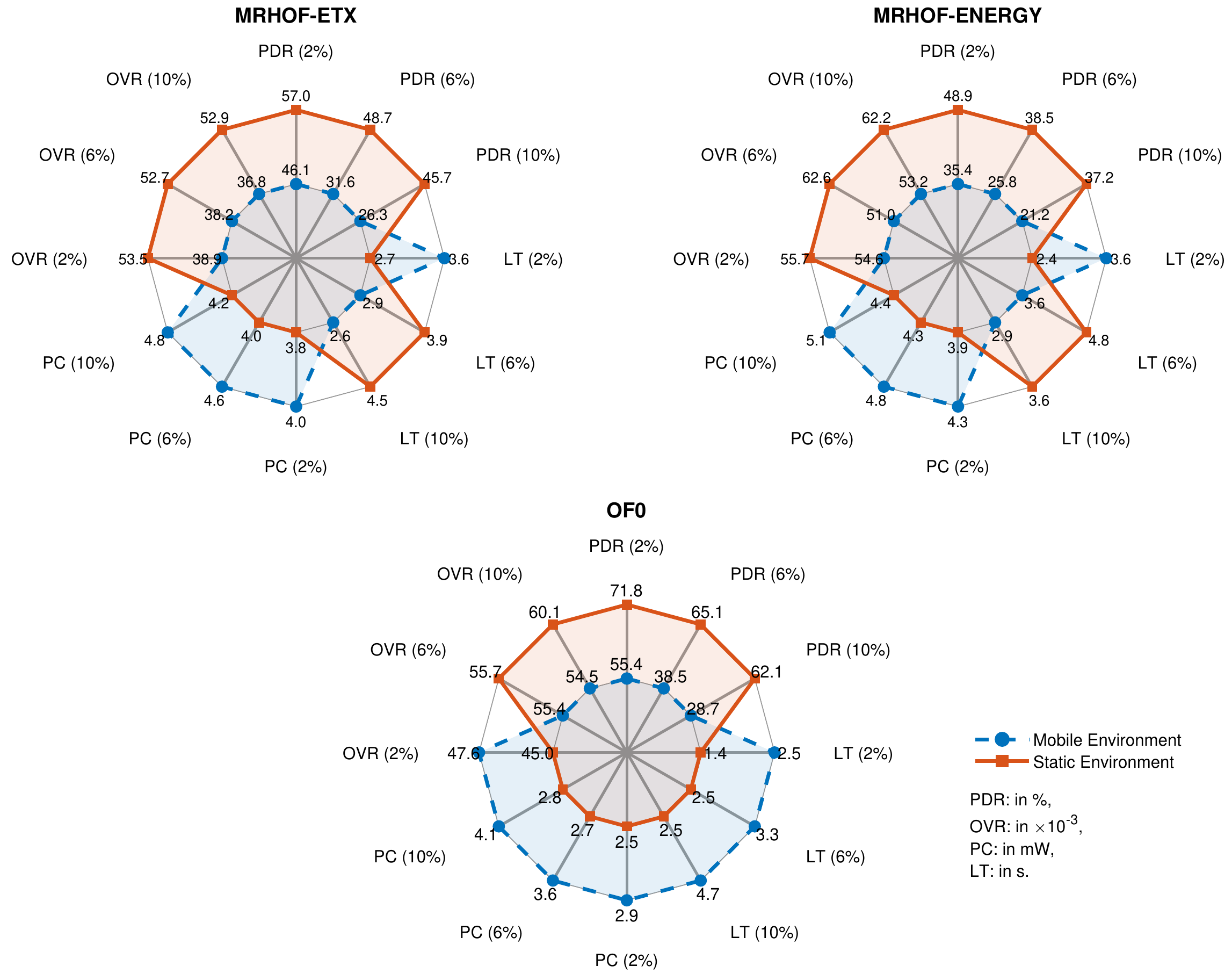}
    \caption{Comparative performances of OFs for version number attack.}
    \label{fig:version_number}
\end{figure}

\begin{figure}
    \centering
    \includegraphics[width=\linewidth]{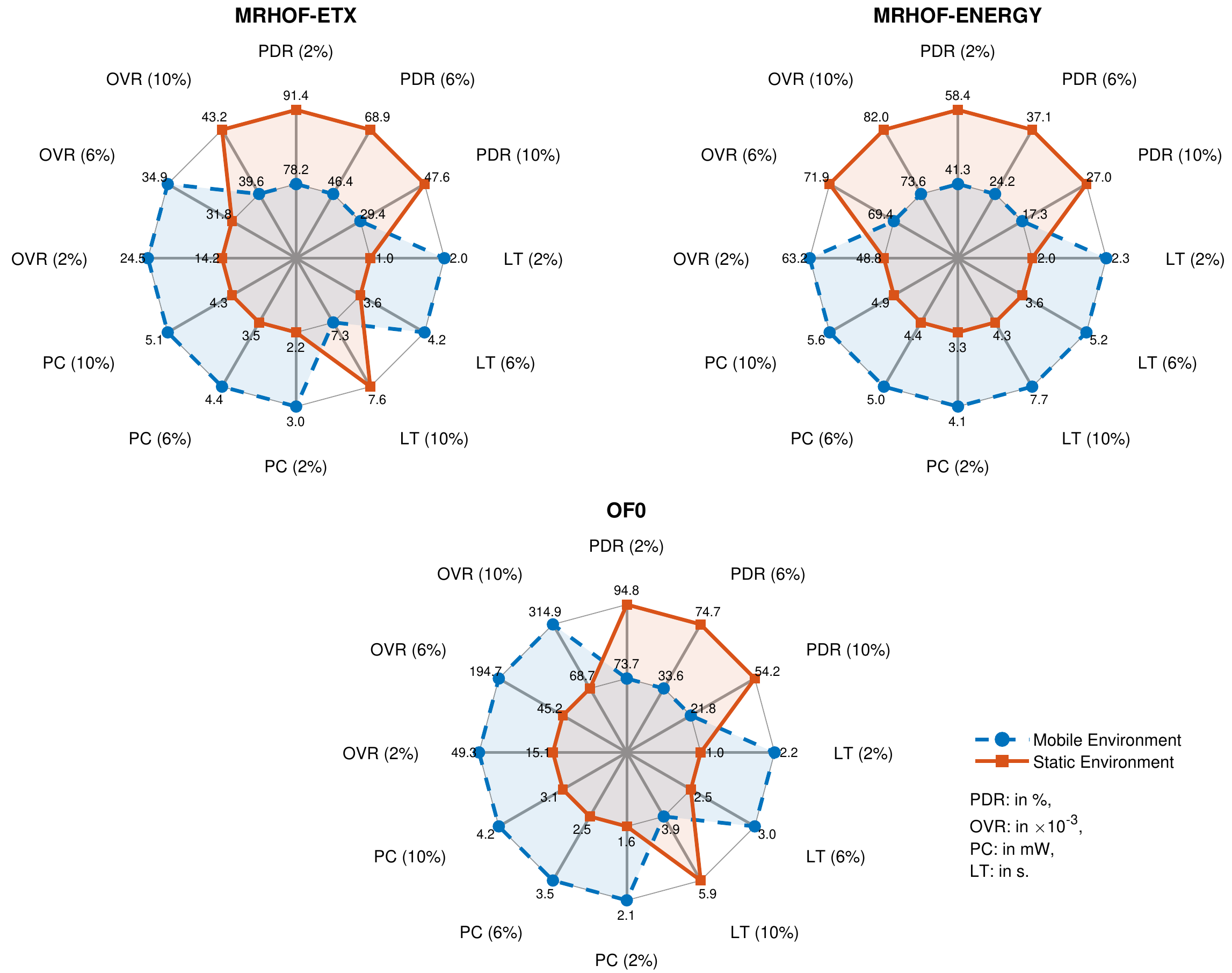}
    \caption{Comparative performances of OFs for DIS flooding attack.}
    \label{fig:dis_flood}
\end{figure}

\begin{figure}
    \centering
    \includegraphics[width=\linewidth]{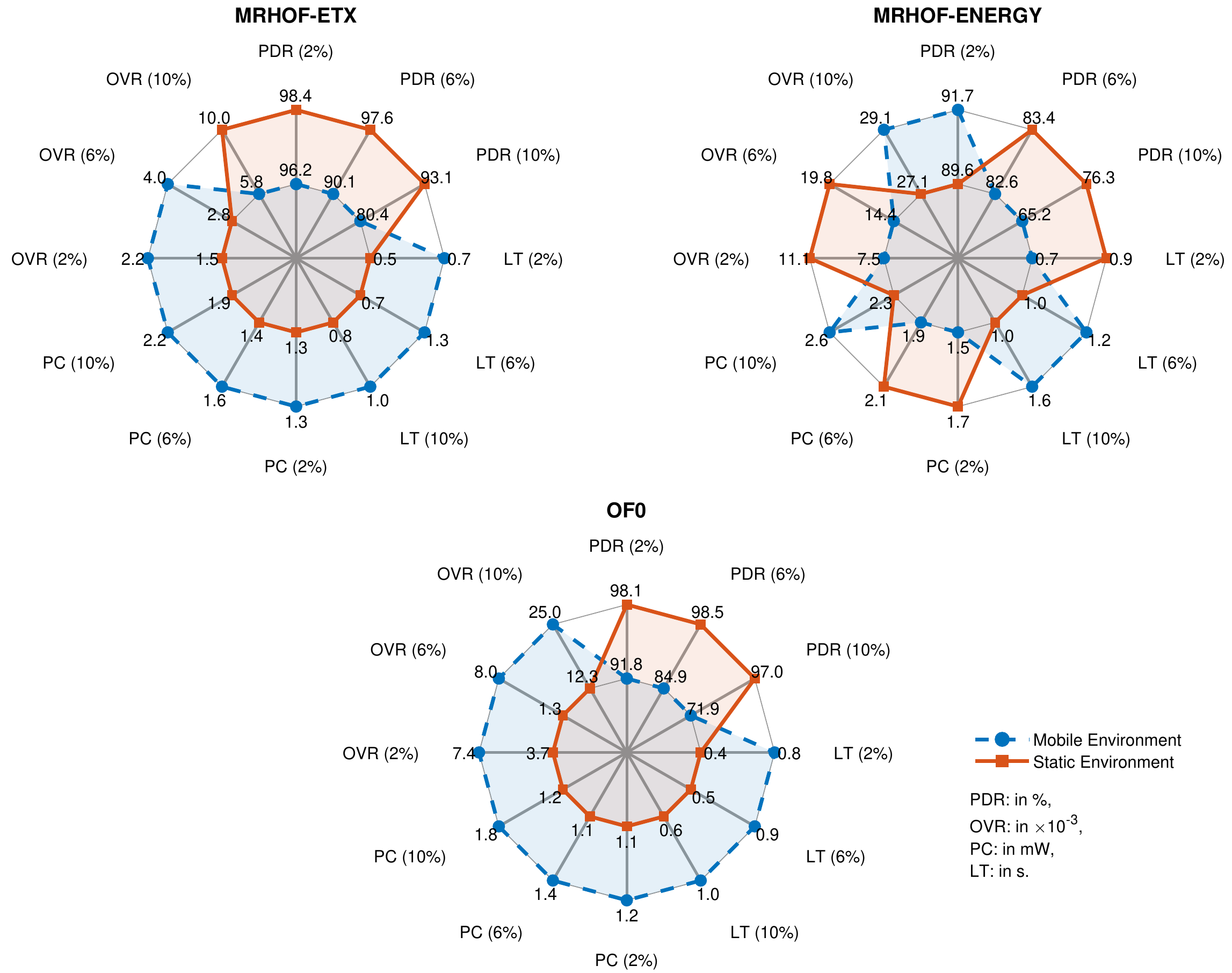}
    \caption{Comparative performances of OFs for worst parent attack.}
    \label{fig:worst_parent}
\end{figure}

The average values of the performance metrics obtained from the ten networks with static and mobile attacker nodes are considered for evaluation. For each OF used (i.e., MRHOF-ETX, MRHOF-ENERGY, and OF0), the comparative results are shown separately for the DIS flooding, the version number, and the worst parent attacks in Figures~\ref{fig:version_number}-\ref{fig:worst_parent}. Note that attacker densities are denoted with the performance metric in the figures. Speaking concretely, `PDR (10\%)' represents average PDR value when 10\% of all nodes are mobile attack nodes in the network.

The PDR values obtained from MRHOF-ETX, MRHOF-ENERGY, and OF0 clearly suggest that the existence of mobile attacker nodes adversely affects the network. Although not significant, an exceptional case is observed only when MRHOF-ENERGY is used when the network is subject to the worst parent attack with an attacker density of 2\%. It is also seen from the figures that the differences in the PDR performances often increase as more mobile attackers are involved in the network. Although MRHOF-ENERGY has shown the least PDR performances on both static and mobile environments, it becomes more resistant (10.20\% performance degradation on average) to the mobility status of attackers than MRHOF-ETX and OF0, while OF0 is the most sensitive (24\% performance degradation on average) to the attacker's mobility status.

Similarly to PDR, the existence of mobile attackers dramatically harms the PC performance of the nodes in the network for all types of attacks, and a positive correlation between the density of the mobile attackers and overall power consumption can also be observed in the figures. The only exception here is observed with MRHOF-ENERGY when the network is subject to the worst parent attack with attacker densities of 2\% and 6\%. The least and highest differences in PC performance are obtained with MRHOF-ENERGY (0.40mW increase on average), OF0 (0.69mW increase on average), respectively. However, no matter the attackers are mobile or static, MRHOF-ENERGY yields the highest PC of the nodes, whereas a lowest PC of the nodes is obtained with OF0 on overall.

Unlike the performance of PDR or PC, a clear correlation between the OVR metric and the mobility status of attackers can hardly be concluded. It is seen from the results that, particularly for the version number attack, a lower OVR value is observed contrarily when the attackers in the network switch from static to mobile. For this attack, it is seen that OVR dramatically reduces when attackers become mobile and when MRHOF-ETX is used. This is because mobile attackers can be out of the coverage area of other nodes in the network when the trickle timers are reset, preventing falsified DIO packets from being received by others. This saves the network from unnecessarily rebuilding the whole DODAG graph through the control packets. For DIS flood attack, it is seen that the average OVR increases as the attackers becomes denser in the network, which is the case observed also in PDR and PC evaluation. As in version number attack, a lower OVR can also be observed on average when attackers become mobile. This is because mobile attackers, although rarely, can reach somewhere outside the coverage area, which prevents them from triggering neighboring nodes to send DIS packets. It should be noted from the results that a dramatic jump in OVR is observed with OF0 for the DIS flooding attack when three or five attackers become mobile in the network. This indicates that OF0 is vulnerable to the increase in the number of mobile attackers. As for the worst parent attack, we can easily conclude that the lowest OVR is obtained with all OFs regardless of the attacker being statically positioned or mobile. 

Similar to OVR performance, higher LT performances can be observed when the network is subject to version number and DIS flooding attacks no matter the attackers are static or mobile. This is due to the additional OVR introduced by these attacks. 
This is because mobile nodes are unable to trigger a global repair in the network as they are likely to be out of coverage of parent nodes most of the time, which prevents the network from being congested. Moreover, an interesting results here is that, particularly for version number attack, much lower LT performance is observed when attackers are static in the network. The least difference in LT performance is observed with MRHOF-ETX on average, while MRHOF-ENERGY and OF0 performs very similar to each other.

In order to reveal how large mobile attackers can down the performance of RPL than static attackers with respect to the OFs, we have thoroughly analyzed the differences in the performance of OFs separately for version number, DIS flooding, and worst parent attacks. The overall differences are given in Table~\ref{tbl:comparative_table}. Note that the performance results, which are obtained when the network is run with static environment, are taken from~\cite{Doan2022AnalysisOR}. The values in this table represent the overall performance differences between when the network is subject to mobile attackers and static attackers, and so they imply how the network worsened when the attackers become mobile. Note that, the biggest gap in the performance are highlighted with gray color in the table, and that the negative values in this table imply that positive effect is observed when the attackers become mobile. 

\begin{table}[!ht]
\caption{Comparative performance differences of OFs for different types of attack.}
\footnotesize
\centering
\begin{tabular}{llcccc} \cline{1-6}
\textbf{Attack}                                                                    & \textbf{OF}     & \multicolumn{1}{l}{\textbf{PDR (in \%)}} & \multicolumn{1}{l}{\textbf{LT (in s)}} & \multicolumn{1}{l}{\textbf{PC (in mW)}} & \multicolumn{1}{l}{\textbf{OVR (in $\times10^{-3}$)}} \\ \cline{1-6}
\multirow{3}{*}{\textbf{\begin{tabular}[c]{@{}l@{}}Version\\ Number\end{tabular}}} & \textbf{ETX}    & 15.83                                    & -0.64                                   & 0.49                                    & -15.08                                   \\
                                                                                   & \textbf{ENERGY} & 14.08                                    & -0.25                                   & 0.54                                    & -7.22                                    \\
                                                                                   & \textbf{OF0}    & \cellcolor{gray75}25.48                                    & \cellcolor{gray75}1.35                                    & \cellcolor{gray75}0.86                                    & \cellcolor{gray75}-1.08                                    \\ \cline{2-6}
\multirow{3}{*}{\textbf{\begin{tabular}[c]{@{}l@{}}DIS\\ Flooding\end{tabular}}}   & \textbf{ETX}    & 17.95                                    & 0.44                                    & 0.84                                    & 3.27                                     \\
                                                                                   & \textbf{ENERGY} & 13.23                                    & \cellcolor{gray75}1.75                                    & 0.66                                    & 1.13                                     \\
                                                                                   & \textbf{OF0}    & \cellcolor{gray75}31.51                                    & -0.12                                   & \cellcolor{gray75}0.91                                    & \cellcolor{gray75}143.29                                   \\ \cline{2-6}
\multirow{3}{*}{\textbf{\begin{tabular}[c]{@{}l@{}}Worst\\ Parent\end{tabular}}}   & \textbf{ETX}    & 7.42                                     & 0.32                                    & 0.21                                    & -0.77                                    \\
                                                                                   & \textbf{ENERGY} & 3.28                                     & 0.21                                    & 0.01                                    & -2.32                                    \\
                                                                                   & \textbf{OF0}    & \cellcolor{gray75}14.99                                    & \cellcolor{gray75}0.37                                    & \cellcolor{gray75}0.31                                    & \cellcolor{gray75}7.65    \\ \cline{1-6}                                
\end{tabular}
\label{tbl:comparative_table}
\end{table}

From the performance differences given in the table, it can easily be concluded that the PDR, LT, PC, and OVR performances of the network down much bigger when the attackers become mobile and when routing is governed OF0 for all types of attack. This finding is not surprising because, as stressed in~\cite{Doan2022AnalysisOR}, OF0 is more robust to static attackers yielding higher network performance than MRHOF. Therefore, much bigger reaction to the mobile attackers can be observed with OF0.
\section{Conclusion}
\label{section_conclusion}
In this study, the performance of RPL is evaluated against mobile attackers. This is the first study that has detailed analysis of OF0, MRHOF-ETX and MRHOF-ENERGY with realistic networks that consist of mobile attackers. We believe that assessing the impact of mobile attackers is a significant step in order to develop security solutions for RPL specific attacks. As a future study, we are planning to explore the effects of attacks when victim nodes are mobile, and develop security solutions for such mobile IoT networks.

\bibliographystyle{unsrt}  
\bibliography{references}

\begin{thebibliography}{10}

\bibitem{Statistica}
Statistica.
\newblock Internet of things (iot) connected devices installed base worldwide
  from 2015 to 2025 (in billions).
\newblock
  \url{https://www.statista.com/statistics/471264/iot-number-of-connected-devices-worldwide/},
  2016.
\newblock Accessed: 2022-06-01.

\bibitem{of_metrics}
Jean-Philippe Vasseur, Mijeom Kim, Kristofer S.~J. Pister, Nicolas Dejean, and
  Dominique Barthel.
\newblock Routing metrics used for path calculation in low-power and lossy
  networks.
\newblock {\em RFC}, 6551:1--30, 2012.

\bibitem{kamgueu2018survey}
Patrick~Olivier Kamgueu, Emmanuel Nataf, and Thomas~Djotio Ndie.
\newblock Survey on rpl enhancements: A focus on topology, security and
  mobility.
\newblock {\em Computer Communications}, 120:10--21, 2018.

\bibitem{lamaazi2018rpl}
Hanane Lamaazi, Nabil Benamar, and Antonio~J Jara.
\newblock Rpl-based networks in static and mobile environment: A performance
  assessment analysis.
\newblock {\em Journal of King Saud University-Computer and Information
  Sciences}, 30(3):320--333, 2018.

\bibitem{le2013impact}
Anhtuan Le, Jonathan Loo, Aboubaker Lasebae, Alexey Vinel, Yue Chen, and
  Michael Chai.
\newblock The impact of rank attack on network topology of routing protocol for
  low-power and lossy networks.
\newblock {\em IEEE Sensors Journal}, 13(10):3685--3692, 2013.

\bibitem{mayzaud2014study}
Anth{\'e}a Mayzaud, Anuj Sehgal, R{\'e}mi Badonnel, Isabelle Chrisment, and
  J{\"u}rgen Sch{\"o}nw{\"a}lder.
\newblock A study of rpl dodag version attacks.
\newblock In {\em IFIP international conference on autonomous infrastructure,
  management and security}, pages 92--104. Springer, 2014.

\bibitem{aris2016rpl}
Ahmet Aris, Sema~F Oktug, and S~Berna~Ors Yalcin.
\newblock Rpl version number attacks: In-depth study.
\newblock In {\em NOMS 2016-2016 IEEE/IFIP Network Operations and Management
  Symposium}, pages 776--779. IEEE, 2016.

\bibitem{Doan2022AnalysisOR}
Cansu Dogan., Selim Yilmaz., and Sevil Sen.
\newblock Analysis of rpl objective functions with security perspective.
\newblock In {\em Proceedings of the 11th International Conference on Sensor
  Networks - SENSORNETS}, pages 71--80. SciTePress, INSTICC, 2022.

\bibitem{thubert2012objective}
Pascal Thubert.
\newblock {Objective Function Zero for the Routing Protocol for Low-Power and
  Lossy Networks (RPL)}.
\newblock RFC 6552, March 2012.

\bibitem{gnawali2012minimum}
Omprakash Gnawali and Philip Levis.
\newblock The minimum rank with hysteresis objective function.
\newblock {\em RFC 6719}, 2012.

\bibitem{winter2012rpl}
Tim Winter, Pascal Thubert, Anders Brandt, Jonathan~W Hui, Richard Kelsey,
  Philip Levis, Kris Pister, Rene Struik, Jean-Philippe Vasseur, Roger~K
  Alexander, et~al.
\newblock Rpl: Ipv6 routing protocol for low-power and lossy networks.
\newblock {\em rfc}, 6550:1--157, 2012.

\bibitem{mayzaud2016taxonomy}
Anth{\'e}a Mayzaud, Remi Badonnel, and Isabelle Chrisment.
\newblock A taxonomy of attacks in rpl-based internet of things.
\newblock {\em I. J. Network Security}, 18:459--473, 2016.

\bibitem{pancaroglu2021load}
Doruk Pancaroglu and Sevil Sen.
\newblock Load balancing for rpl-based internet of things: A review.
\newblock {\em Ad Hoc Networks}, page 102491, 2021.

\bibitem{Zaatouri_et_al}
Ibtissem Zaatouri, Francoise Sailhan, Stephane Rovedakis, Awatef Guiloufi,
  Nouha Alyaoui, and Abdennaceur Kachouri.
\newblock Performance evaluation of mobility-aware routing protocol for low
  power and lossy networks.
\newblock In {\em 2019 16th International Multi-Conference on Systems, Signals
  \& Devices (SSD)}, pages 636--641, 2019.

\bibitem{kim2017challenging}
Hyung-Sin Kim, Jeonggil Ko, David~E Culler, and Jeongyeup Paek.
\newblock Challenging the ipv6 routing protocol for low-power and lossy
  networks (rpl): A survey.
\newblock {\em IEEE Communications Surveys \& Tutorials}, 19(4):2502--2525,
  2017.

\bibitem{Cotrim_et_al}
Jeferson Rodrigues~Cotrim and Joao~Henrique Kleinschmidt.
\newblock Performance evaluation of rpl on a mobile scenario with different
  contikimac radio duty cycles.
\newblock In {\em 2017 IEEE 18th International Conference on High Performance
  Switching and Routing (HPSR)}, pages 1--6, 2017.

\bibitem{RPL_under_mobility}
Kevin~C. Lee, Raghuram Sudhaakar, Lillian Dai, Sateesh Addepalli, and Mario
  Gerla.
\newblock Rpl under mobility.
\newblock In {\em 2012 IEEE Consumer Communications and Networking Conference
  (CCNC)}, pages 300--304, 2012.

\bibitem{reverse_trickle_timer_1}
Cosmin Cobârzan, Julien Montavont, and Thomas Noël.
\newblock Analysis and performance evaluation of rpl under mobility.
\newblock In {\em 2014 IEEE Symposium on Computers and Communications (ISCC)},
  pages 1--6, 2014.

\bibitem{reverse_trickle_timer_2}
Harith Kharrufa, Hayder Al-Kashoash, Yaarob Al-Nidawi, Maria~Quezada Mosquera,
  and A.H. Kemp.
\newblock Dynamic rpl for multi-hop routing in iot applications.
\newblock In {\em 2017 13th Annual Conference on Wireless On-demand Network
  Systems and Services (WONS)}, pages 100--103, 2017.

\bibitem{Murali_and_Jamalipour}
Sarumathi Murali and Abbas Jamalipour.
\newblock Mobility-aware energy-efficient parent selection algorithm for low
  power and lossy networks.
\newblock {\em IEEE Internet of Things Journal}, 6(2):2593--2601, 2019.

\bibitem{Co-rpl}
Olfa Gaddour, Anis Koubäa, Raghuraman Rangarajan, Omar Cheikhrouhou, Eduardo
  Tovar, and Mohamed Abid.
\newblock Co-rpl: Rpl routing for mobile low power wireless sensor networks
  using corona mechanism.
\newblock In {\em Proceedings of the 9th IEEE International Symposium on
  Industrial Embedded Systems (SIES 2014)}, pages 200--209, 2014.

\bibitem{Thulasiraman_and_Wang}
Preetha Thulasiraman and Yizhong Wang.
\newblock A lightweight trust-based security architecture for rpl in mobile iot
  networks.
\newblock In {\em 2019 16th IEEE Annual Consumer Communications Networking
  Conference (CCNC)}, pages 1--6, 2019.

\bibitem{Murali_and_Jamalipour_ABC}
Sarumathi Murali and Abbas Jamalipour.
\newblock A lightweight intrusion detection for sybil attack under mobile rpl
  in the internet of things.
\newblock {\em IEEE Internet of Things Journal}, 7(1):379--388, 2020.

\bibitem{ABC}
Dervis Karaboga and Bahriye Basturk.
\newblock A powerful and efficient algorithm for numerical function
  optimization: artificial bee colony (abc) algorithm.
\newblock {\em Journal of Global Optimization}, 39(3):459--471, Nov 2007.

\bibitem{verma2020mitigation}
Abhishek Verma and Virender Ranga.
\newblock Mitigation of dis flooding attacks in rpl-based 6lowpan networks.
\newblock {\em Transactions on emerging telecommunications technologies},
  31(2):e3802, 2020.

\bibitem{aschenbruck2013mobility}
N~Aschenbruck, R~Ernst, E~Gerhards-Padilla, and M~BonnMotion Schwamborn.
\newblock A mobility scenario generation and analysis tool.
\newblock {\em Osnabruck, Germany}, 2013.

\bibitem{osterlind2006cross}
Fredrik Osterlind, Adam Dunkels, Joakim Eriksson, Niclas Finne, and Thiemo
  Voigt.
\newblock Cross-level sensor network simulation with cooja.
\newblock In {\em Proceedings. 2006 31st IEEE Conference on Local Computer
  Networks}, pages 641--648. IEEE, 2006.

\bibitem{contiki-ng}
Contiki-Ng.
\newblock contiki-ng/contiki-ng, 2004.
\newblock [accessed 13-July-2021].

\bibitem{dunkels2011powertrace}
Adam Dunkels, Joakim Eriksson, Niclas Finne, and Nicolas Tsiftes.
\newblock Powertrace: Network-level power profiling for low-power wireless
  networks, 2011.

\end{thebibliography}

\end{document}